\documentclass{new_tlp}
\usepackage{times}
\usepackage{graphicx}
\usepackage{latexsym}
\usepackage{helvet}
\usepackage{courier}
\usepackage{amsmath}
\usepackage{amssymb}
\usepackage{algorithm}
\usepackage{alltt}
\usepackage{verbatim}
\usepackage{url}
\usepackage{xspace}
\usepackage[defblank]{paralist}

\usepackage{pgfplots}
\pgfplotsset{compat=1.14}
\usepackage{tikz}
\usepackage[caption=false]{subfig}

\usepackage{algorithmicx}
\usepackage{algpseudocode}

\usepackage{pgf}
\usepackage{tikz}
\usetikzlibrary{arrows,automata}
\usepackage{inputenc}
\usepackage{verbatim}

\newcommand{\ignore}[1]{}
\newcommand{\finish}[1]{}
\newcommand{\skipit}[1]{{ #1}}

\newcommand{\DL}{{ DL}}

\newcommand{\pl}{\partial_{||}}
\newcommand{\md}[1]{-\partial #1}

\newcommand{\mt}[1]{\mathtt{#1}}

\newcommand{\true}{{\bf true}}
\newcommand{\false}{{\bf false}}

\newcommand{\unknown}{{\bf unknown}}
\renewcommand{\imath}{i}

\newcommand{\Zp}{\Z^{+1}}
\newcommand{\Zm}{\Z^{-1}}
\newcommand{\UU}{\U\U}

\newcommand{\Datalog}{Datalog$^\neg$}

\newcounter{clause}
\def\theclause{$c$\arabic{clause}}

\newenvironment{clause}{\begin{tabbing}
xxx\=xxx\=xxx\=\+\kill}%
{\end{tabbing}}

\newenvironment{Clause}{\refstepcounter{clause}%
\begin{tabbing}
cxxxx\=xxx\=xxx\=\kill
\theclause\>\+}%
{\end{tabbing}}

\newtheorem{theorem}{Theorem}
\newtheorem{lemma}[theorem]{Lemma}
\newtheorem{definition}[theorem]{Definition}
\newtheorem{propn}[theorem]{Proposition}
\newtheorem{proposition}[theorem]{Proposition}
\newtheorem{corollary}[theorem]{Corollary}
\newtheorem{example}[theorem]{Example}

\newcommand{\F}{{\cal F}}
\newcommand{\G}{{\cal G}}

\renewcommand{\P}{{\cal P}}
\newcommand{\Q}{{\cal Q}}

\newcommand{\U}{{\cal U}}

\newcommand{\V}{{\cal V}}
\newcommand{\W}{{\cal W}}

\newcommand{\Z}{{\cal Z}}

\ignore{
\newcounter{clause}
\def\theclause{$c$\arabic{clause}}

{\end{tabbing}}

{\end{tabbing}}

\newtheorem{theorem}{Theorem}
\newtheorem{lemma}[theorem]{Lemma}

\newtheorem{propn}[theorem]{Proposition}
\newtheorem{proposition}[theorem]{Proposition}

\newenvironment{proof}[1][Proof]{\begin{trivlist}\item[\hskip \labelsep {\bfseries #1}]}{\end{trivlist}}
}

\def    \md      {\sqsupseteq}
\def    \len     { \leq_{ -1 } \, }

\def    \gen     { \geq_{ -1 } \, }

\def    \lep     { \leq_{ +1 } \, }

\def    \gep     { \geq_{ +1 } \, }

\def    \mdn     { \md_{ -1 } \, }
\def    \mdp     { \md_{ +1 } \, }

\def    \lpm     { \leq_{ \pm } \, }

\begin{document}

\title{On Signings and the Well-Founded Semantics}

\author[M.J. Maher]{Michael J. Maher \\
Reasoning Research Institute \\
Canberra, Australia  \\
\email{michael.maher@reasoning.org.au}
}


\maketitle
\bibliographystyle{acmtrans}

\begin{abstract}
We use Kunen's notion of a signing to establish two theorems about 
the well-founded semantics of logic programs,
in the case where we are interested in only (say) the positive literals of a predicate $p$
that are consequences of the program.
The first theorem identifies a class of programs for which the well-founded and Fitting semantics
coincide for the positive part of $p$.
The second theorem shows that if a program has a signing then
computing the positive part of $p$ under the well-founded semantics
requires the computation of only one part of each predicate.
This theorem suggests an analysis for query-answering under the well-founded semantics.
In the process of proving these results,
we use an alternative formulation of the well-founded semantics of logic programs,
which might be of independent interest.
\end{abstract}


\section{Introduction}

A signing of a logic program identifies a structure of the program
that ensures that no predicate depends both positively and negatively on another predicate.
Signings have been used in several previous works.
\cite{Kunen89} introduced signings and used them to establish cases where
a three-valued model of the Clark-completion of $P$ can be extended to a two-valued model.
Any program has a corresponding signed program that is equivalent wrt SLDNF-resolution \cite{strictcomp}.
\cite{Dung92} uses signings to establish existence of a stable model
and to identify cases where the well-founded and sceptical stable model semantics coincide.
\cite{Turner93}  shows that signed logic programs have at least two stable models,
unless the well-founded model is itself a stable model.

In this note we address the situation where we are only interested in (say)
the positive consequences of a program, for a predicate $p$ (which we term the \emph{positive part} of $p$).
This situation arose in the implementation of a defeasible logic \cite{sdl}
by compilation to \Datalog{} \cite{sdl2}.
Using signings, we identify a case where the well-founded and Fitting semantics
coincide for the positive part of $p$.
More importantly, we show that, for programs with a signing, the positive part of $p$
can be computed using only one part (positive or negative) of each predicate it depends on.
This result extends to an analysis that can determine the extent to which a program
without a signing can avoid computing parts of predicates.
The proofs are facilitated by an alternative formulation of the well-founded semantics,
which might be of independent interest.

The paper is structured as follows.
In the next section, we provide the background needed,
including defining signings and the well-founded semantics.
Following that, in Section \ref{sec:alt}, we define an alternative formulation of the well-founded semantics.
In Section \ref{sec:results} we establish the two theorems.
The  appendix contains re-proofs of the results by Kunen, Dung, and Turner.

\section{Preliminaries}  \label{sec:LP}

We introduce the elements of logic programming that we will need.
The first part defines notation and terminology for syntactic aspects of logic programs,
including signings.
The second part defines the semantics of logic programs we will use.

\subsection{Syntax and Structure of Logic Programs} \label{sec:syntax}

Let $\Pi$ be a set of predicate symbols, $\Sigma$ be a set of function symbols,
and $\V$ be a set of variables.
Each symbol has an associated arity greater or equal to 0.
A function symbol of arity 0 is called a \emph{constant},
while a predicate of arity 0 is called a \emph{proposition}.
The \emph{terms} are constructed inductively in the usual way:
any variable or constant is a term;
if $f \in \Sigma$ has arity $n$ and $t_1, \ldots, t_n$ are terms then $f(t_1, \ldots, t_n)$ is a term;
all terms can be constructed in this way.
An \emph{atom} is constructed by applying a predicate $p \in \Pi$ of arity $n$ to $n$ terms.
A \emph{literal} is either an atom or a negated atom $not~ a$, where $a$ is an atom.

A {\em logic program} 
is a collection of {\em clauses} of the form
$$a ~\mbox{:-}~ b_1 , \ldots , b_m ,
not~ c_1 , \ldots , not~ c_n$$
where 
$a, b_1 , \ldots , b_m , c_1 , \ldots , c_n$
are atoms ($m \geq 0 , n \geq 0$).
The positive literals and the negative literals
are grouped separately
purely for notational convenience.
$a$ is called the \emph{head} of the clause and the remaining literals form the \emph{body}.
Sometimes, for brevity, we write a rule as $a ~\mbox{:-}~ B$.
The set of all clauses with predicate symbol $p$ in the head
are said to be the clauses defining $p$.
We use \emph{ground} as a synonym for variable-free.
The set of all variable-free instances of clauses in a logic program $P$
is denoted by $ground(P)$.
$ground(P)$ can be considered a propositional logic program but it is infinite, in general.
For the semantics we are interested in, $P$ and $ground(P)$ are equivalent
with respect to inference of ground literals.

We present some notions of dependence among predicates that are derived purely from
the syntactic structure of a logic program $P$.
We follow the notation and definitions of \cite{Kunen89}.
$p$, $q$ and $r$ range over predicates.
We define
$p \mdp q$
if $p$ appears in the head of a rule and $q$
is the predicate of a positive literal in the body of that rule.
$p \mdn q$
if $p$ appears in the head of a rule and $q$
is  the predicate of a negative literal in the body of that rule.
%

We  can represent these dependencies of $P$ in a \emph{signed predicate dependency graph} $\G$
where  each predicate forms a vertex and there is an edge from $p$ to $q$ labeled $+$ iff $p \mdp q$
and an edge from $p$ to $q$ labeled $-$ iff $p \mdn q$.
(See Figure~\ref{fig:pdg} for an example.)
Given a logic program, this graph can be constructed in linear time.

Building on these dependencies, we can define further dependencies and properties of a program.
We define $p \md q$ iff $p \mdp q$ or $p \mdn q$,
and say $p$ \emph{directly depends} on $q$.
The transitive closure of $\md$ is denoted by $\ge$.
If $p \ge q$, we say $p$ \emph{depends} on $q$.
We define
$p \approx q$ iff $p \geq q$ and $q \geq p$, expressing that $p$ and $q$ are mutually recursive.
%
%
The binary relations
$\gep$ and $\gen$ are defined inductively as  the least relations such that,
for all predicates $p$, $q$ and $r$,
\[p \gep p\]
and
\[p \md_i q ~ \mbox{ \em and } ~
q \geq_j r  ~ \mbox{ \em implies } ~
p \geq_{ i \cdot j } r\]
where $i \cdot j$ denotes multiplication of $i$ and $j$.\footnote{
Equivalently, if we view these conditions as definite clauses, $\gep$ and $\gen$
are determined by the least model of these clauses and facts for $\mdp$ and $\mdn$.
}
Essentially, $\gep$ denotes a relation of
dependence through an even number of negations
and $\gen$ denotes
dependence through an odd number of negations.
As is usual, we will write $p \leq q$ when $q \geq p$,
and similarly for other orderings.
The transitive closure of $\mdp$ is denoted by $\geq_0$ \cite{Fages94}\footnote{
Note that, despite the notation, $\geq$ and $\geq_0$ are not necessarily reflexive:
if $p$ does not depend on $p$, then we do not have $p \geq p$ nor $p \geq_0 p$.
}.
$P$ is \emph{positive order-consistent} \cite{Fages94} (also referred to as $P$ being \emph{tight})
if $\geq_0$ is well-founded on $ground(P)$.

A program $P$ is \emph{stratified} if there are no predicates $p$ and $q$ in $\Pi$ such that
$p \approx q$ and $p \gen q$.
In other words, $P$ is stratified if there is no negative dependency within a strongly connected component
of the signed predicate dependency graph.
This is equivalent to the original definition \cite{ABW} that explicitly identified strata.
A program $P$ is said to be \emph{strict} if no predicate $p$ depends both positively and negatively
on a predicate $q$, that is, we never have $p \gep q$ and $p \gen q$.

A set $\P \subseteq \Pi$ of predicates in a program $P$ is \emph{downward-closed} if,
whenever $p \in \P$ and $q \leq p$, then $q \in \P$.
$\P$ is  \emph{downward-closed with floor} $\F$
if $\F \subseteq \P$ and both $\P$ and $\F$ are downward-closed.
The notion of downward-closed with floor underlies
Dung's notion of bottom-stratified and top-strict logic programs \cite{Dung92}.
Indeed, if $\P$ is downward-closed with floor $\F$,
$P$ is strict on $\P\backslash\F$,
and $P$ is stratified on $\F$,
then $P$ is bottom-stratified and top-strict.

A \emph{signing} on $\P$ for a program $P$ is a function $s$ that maps $\P$ to $\{-1, +1\}$
such that, for $p, q \in \P$, $p \leq_i q$ implies $s(p) = s(q) \cdot i$.
A signing is extended to atoms by defining $s( p(\vec{a}) ) = s(p)$.
For any signing $s$ for a set of predicates $\P$, there is an inverted signing $\bar{s}$, defined by
$\bar{s}(p) = - s(p)$.
$s$ and $\bar{s}$ are equivalent in the sense that they partition $\P$ into the same two sets.
Obviously, $\bar{\bar{s}} = s$.
Let all the predicates of $P$ be contained in $\P$.
If $\P$ has a signing for $P$, then $P$ is strict;
if $P$ has a $\geq$-largest predicate and $P$ is strict, then $\P$ has a signing for $P$ \cite{Kunen89}.

It is relatively easy to determine whether a program $P$ has a signing on $\P$.
We assume an undirected version of the signed predicate dependency graph $\G$, restricted to $\P$.
We analyse each connected component of $\G$ separately.
Arbitrarily choose a vertex in the component and assign it a value in $\{+1, -1\}$.
\footnote{
If it is desired that a specific predicate has a positive sign, that can be enforced here.
}
We now propagate signs to adjacent vertices: if $p$ has sign $i$ and there is an (undirected) edge to $q$ of sign $j$
(that is, either $p \md_j q$ or $q \md_j p$),
then $q$ must have sign $i \cdot j$, if there is a signing.
Thus, if such a vertex $q$ is unassigned, it is assigned $i \cdot j$.
If it already has a different sign than $i \cdot j$, then we halt with failure:
there is no signing.
Otherwise, iterating this process for each newly-signed vertex generates a signing for the component.
The whole propagation, for all components, can be done during a single depth-first search of $\G$.
Thus, the computational cost is linear in the size of $\G$ \cite{Cormen} and, hence, linear in the size of $P$.

\begin{proposition}
Deciding whether a logic program has a signing is computable in linear time.
\end{proposition}

\ignore{
The results in this note do not require any ordering to be well-founded,
not even $\leq^+ \cap \leq^-$.
}

A similar set of dependencies over ground atoms can be defined by
applying these definitions to $ground(P)$,
where ground atoms are considered propositions (i.e. predicates).
The results of this paper continue to hold using those notions.

Given a program $P$,
an infinite sequence of atoms $\{ q_i(\vec{a_i}) \}$ is \emph{unfounded} wrt a set of predicates $\Q$ if, 
for every $i$, $q_i \mdp q_{i+1}$ and $q_i \in \Q$.
We say a predicate $p$ \emph{avoids negative unfoundedness} 
\footnote{
We refrain from referring to this property as  \emph{negative positive order-consistency}.}
wrt a signing $s$ on $\Q$ if
for every negatively signed predicate $q$ on which $p$ depends,
no $q$-atom starts an unfounded sequence wrt $\Q$.

We have the following sufficient condition for avoiding negative unfoundedness.
Let $\geq_0^\Q$ denote the transitive closure of $\mdp$ on $\Q$.
We say a predicate $p$ \emph{avoids negative predicate unfoundedness} 
wrt a signing $s$ on $\Q$ if,
for all $q \leq p$, $ (q \geq_0^\Q q)  \rightarrow s(q) = +1$.
That is, the only predicates on which $p$ depends that take part in a positive cycle among predicates of $\Q$
have a positive sign.

\begin{proposition}   \label{prop:negunf}
Let $P$ be a program, $\Q $ a finite subset of $\Pi$, and $p\in\Q$.
Let $s$ be a signing on $\Q$ for $P$.
If $p$ avoids negative predicate unfoundedness wrt $s$ on $\Q$, then 
$p$ avoids negative unfoundedness wrt $s$ on $\Q$.
\end{proposition}
\skipit{
\begin{proof}
We prove the contrapositive.
Suppose $p$ does not avoid negative unfoundedness.
Then there is a predicate $q$ such that $p \geq q$, $s(q) = -1$, and
$\{ q_i(\vec{a_i}) \}$ forms an unfounded sequence wrt $\Q$ with $q \equiv q_1$.
Then we must have $q \mdp q_1 \mdp q_2 \mdp \cdots$ and, 
because $s(q) = -1$ and the dependencies are positive, $s(q_i)=-1$, for every $i$.
Since $\Q$ has only finitely many predicates, a predicate $q_j$ must be repeated.
Hence, $q_j \geq_0^\Q q_j$ and $s(q_j)=-1$.
Therefore, $p$ does not avoid negative predicate unfoundedness.
\end{proof}
}


Using the signed predicate dependency graph,
we can determine which predicates avoid negative predicate unfoundedness wrt $s$ on $\Q$.
We restrict our attention to the subgraph with vertices in $\Q$.
First, using only the positive dependencies $\mdp$ in the subgraph we find the strongly connected components (SCCs).
Given the signing, all vertices in an SCC have the same sign.
Those SCCs consisting of a single vertex without an edge to itself can be ignored;
of the remaining SCCs, we choose a predicate from each SCC that has a negative sign. 
Then, using the full signed predicate dependency graph,
we mark all vertices that depend on any of the chosen vertices;
these predicates do \emph{not} avoid negative predicate unfoundedness.
Unmarked predicates do avoid negative predicate unfoundedness and, hence,
avoid negative unfoundedness.
This method has a linear cost in the size of the signed predicate dependency graph.

\subsection{Semantics of Logic Programs}

We define the semantics of interest in this paper
and identify important relationships between them.

We will only be interested in Herbrand interpretations and models.
A {\em 3-valued Herbrand interpretation}
is a mapping from ground atoms to one of three truth values:
$\true$, $\false$, and $\unknown$.
This mapping can be extended to all formulas using Kleene's 3-valued logic \cite{Kleene}.
Equivalently, a 3-valued Herbrand interpretation $I$ can be represented as the set of literals
$\{ a ~|~ I(a) = \true \} \cup \{ not~a ~|~ I(a) = \false \}$.
This representation is used in the following definitions.
The interpretations are ordered by the subset ordering on this representation.
Given an interpretation $I$, 
the \emph{positive part} of a predicate $p$ is $\{ p(\vec{a}) ~|~ p(\vec{a}) \in I \}$, the restriction of $I$ to $p$-atoms,
and the \emph{negative part} of $p$ is $\{ not~ p(\vec{a}) ~|~ not~ p(\vec{a}) \in I \}$, 
the restriction of $I$ to negative $p$-literals.

We review some notions of fixedpoints of functions on 3-valued Herbrand interpretations.
An interpretation $X$ is a \emph{pre-fixedpoint} of a function $f$ if $f(X) \subseteq X$;
thus, a pre-fixedpoint is closed under the action of $f$.
$X$ is a \emph{fixedpoint} of $f$ if $f(X) = X$.
$f$ is \emph{monotonic} if $X \leq Y$ implies $f(X) \leq f(Y)$.
For any $X$ and monotonic $f$,
we define $lfp(f, X)$ to be the least fixedpoint greater than (or equal to) $X$.
$lfp(f)$ is the least fixedpoint greater than the bottom element $\emptyset$. 
We refine the $\uparrow$ notation and Kleene sequence.
The \emph{Kleene sequence for $f$ from $X$} is a possibly transfinite sequence of interpretations, 
defined as follows: 

\hspace{1.0cm} $f \uparrow 0 = X$ 

\hspace{1.0cm} $f \uparrow (\beta+1) = f( f \uparrow \beta ) \cup f \uparrow \beta$ 

\hspace{1.0cm} $f \uparrow \alpha = \bigcup_{\beta < \alpha} f \uparrow \beta$ if $\alpha$ is a limit ordinal 

\noindent
If $f$ is monotonic and $lfp(f, X)$ exists then the limit of this sequence is $lfp(f, X)$.
This is different from the usual definition of Kleene sequence, so that it can start from any $X$.
It is straightforward to show that if $X \subseteq f(X)$, then $lfp(f, X)$ exists, by induction on the Kleene sequence,
and that $lfp(f, X)$ is also the least pre-fixedpoint greater than $X$.

Fitting \cite{Fitting} defined a semantics for a logic program $P$ in terms of a function $\Phi_P$
mapping 3-valued interpretations, which we define as follows.
\[
\begin{array}{rcl}
\Phi_P(I) &= & \Phi^+_P(I) \cup \neg~ \Phi^-_P(I) \\

\Phi^+_P(I) &= &  \{a ~|~ \mbox{there is a rule } a \mbox{~:-~} B \mbox{ in } ground(P)
\mbox{ where } I (B) = \true \} \\

\Phi^-_P(I) &= & \{ a ~|~ \mbox{for every rule } a \mbox{~:-~} B \mbox{ in } ground(P)
\mbox{ with head } a, I (B) = \false \}
\end{array}
\]
where $\neg S$ denotes the set $\{ not~s ~|~ s \in S \}$.

\emph{Fitting's semantics} associates with $P$ the least fixedpoint of $\Phi_P$,
that is $lfp(\Phi_P)$.
This is the least 3-valued Herbrand model of the Clark completion $P^*$ of $P$.
Thus, the conclusions justified under this semantics are those formulas that evaluate to $\true$ under all
3-valued Herbrand models of $P^*$.

The \emph{stratified semantics} \cite{ABW} 
(or \emph{iterated fixedpoint} semantics)
applies only when $P$ is stratified.
It is defined in stages, by building up partial models, based on the strata, until a full model is constructed.
For a more complete description, see \cite{ABW,AptBol}.
The stratified semantics extends Fitting's semantics, when the program is stratified.

The well-founded semantics \cite{WF91} extends Fitting's semantics by, roughly,
considering atoms to be false if they are supported only by a ``loop'' of atoms.
This is based on the notion of unfounded sets.

Given a logic program $P$
and a 3-valued interpretation $I$, 
a set $A$ of ground atoms is an \emph{unfounded set
with respect to} $I$ iff each atom $a\in A$ satisfies the
following condition: For each rule $r$ of $ground(P)$
whose head is $a$, (at least) one of the following holds:
\begin{enumerate}
 \item Some literal in the body evaluates to $\false$ in $I$.
 \item Some atom in the body occurs in $A$ 
\end{enumerate}

The \emph{greatest unfounded set} of $P$ with respect to
$I$ (denoted $\U_{P}(I)$) is the union of all the unfounded sets with respect to $I$.
Notice that, if we ignore the second part of the definition of unfounded set wrt $I$,
the definition of unfounded set is the same as the expression inside the definition of $\Phi^-_P(I)$.
It follows that $\Phi^-_P(I) \subseteq \U_{P}(I)$, for every $I$.

The function $\W_{P}(I)$ is  defined by
$\W_{P}=\Phi^+_P(I)  \cup \neg ~ \U_{P}(I)$.
The \emph{well-founded semantics} of a program $P$ is represented by the least fixedpoint of $\W_{P}$.
This is a 3-valued Herbrand model of $P^*$.
Because $\Phi^-_P(I) \subseteq \U_{P}(I)$, for every $I$, we have $\Phi_P(I) \subseteq \W_{P}(I)$, for every $I$,
and, hence, $lfp(\Phi_P) \subseteq lfp(\W_P)$.
That is, Fitting's semantics is weaker than the well-founded semantics.
\cite{HW02} show that the well-founded semantics and the Fitting semantics
are characterized by different partial level mappings, where the level mappings for the well-founded semantics
are more permissive than those for Fitting's semantics.
If $P$ is stratified then the well-founded semantics is a 2-valued Herbrand model of $P^*$ 
and equal to the stratified semantics.

\section{Alternative formulation of the well-founded semantics}   \label{sec:alt}

We will find it useful to have a different formulation of the well-founded semantics
that clarifies how the well-founded semantics extends Fitting's semantics.
We say a set $A$ of ground atoms is a \emph{circular\footnote{
``circular'' is something of a misnomer.  If $ground(P)$ is infinite, then an infinite set $A$ can satisfy the conditions without any circularity.  However, for finite \Datalog{} programs, $A$ does exhibit a cycle.
} unfounded set  with respect to} $I$ iff 
$A$ is an unfounded set wrt $I$ and,
for each $a \in A$ there is a rule $r \in ground(P)$ with head $a$ with no literal evaluating to $\false$ in $I$
(and, hence, some atom $b$ in the body of $r$ occurs in $A$).
The \emph{greatest circular unfounded set} of $P$ with respect to
$I$ (denoted $\Z_{P}(I)$) is the union of all the circular unfounded sets with respect to $I$.
\footnote{The definition of circular unfounded set is similar to the definition in \cite{trans_BU} of an unfounded set
once the program is simplified by all literals inferred by Fitting's semantics,
but here the program is not simplified and, in general, most literals have not been inferred.
}
Obviously, any element of $\Z_{P}(I)$, for any $I$, is the start of an unfounded sequence.
Consequently, if $P$ is positive order-consistent (or tight), then $\Z_{P}(I)$ is empty.

A \emph{minimalistic} circular unfounded set of $P$ wrt $I$
is a circular unfounded set  $S$ of $P$ wrt $I$
such that either it is empty, or it cannot be partitioned into two disjoint nonempty circular unfounded sets\footnote{
A minimalistic circular unfounded set is not necessarily a minimal circular unfounded set.
For example, given the program $\{ q \mbox{\,:-\,} p. ~~ p \mbox{\,:-\,} p. \}$,
the set $\{ p, q \}$ is minimalistic, because it cannot be partitioned into disjoint nonempty circular unfounded sets,
but not minimal because $\{ p \}$ is a circular unfounded set.
}.
If $P$ has a signing $s$, all elements of a minimalistic circular unfounded set must have the same sign.
Thus, with a slight abuse of language, we can say that minimalistic circular unfounded sets have a sign.
Let a program $P$ and a signing $s$ for $P$ be fixed.
We define $\Zm_P(I)$ to be the greatest circular unfounded set  of $P$ wrt $I$ 
consisting only of atoms $a$ with $s(a) = -1$.
It can be obtained as the union of all minimalistic circular unfounded sets with sign $-1$.
Similarly, $\Zp_P(I)$ is the greatest circular unfounded set with sign $+1$.
It should be clear that $\Z_P(I) = \Zp_P(I) \cup \Zm_P(I)$.

Define $\W'_{P}(I)=\Phi^+_P(I)  \cup \neg ~ \Phi^-_P(I) \cup \neg ~ \Z_{P}(I)$.
Then $\W'_{P} (I) \subseteq \W_{P}(I)$, for every $I$,
since both $\Phi^-_P(I)$ and $\Z_{P}(I)$ are clearly subsets of $\U_{P}(I)$.
We now show that $W_P$ and $W'_P$ have the same pre-fixedpoints and fixedpoints.

\begin{lemma}   \label{lemma:prefp}
For every 3-valued interpretation $I$,
\[
\W_P(I) \subseteq I  ~~~\mbox{    iff    }~~~  \W'_P(I) \subseteq I 
\]
and
\[
\W_P(I) = I  ~~~\mbox{    iff    }~~~  \W'_P(I) = I 
\]
\end{lemma}
\skipit{
\begin{proof}
Part 1.
One direction is straightforward:
since $\W'_{P} (I) \subseteq \W_{P}(I)$,
if $\W_P(I) \subseteq I$ then also $\W'_P(I) \subseteq I$.

For the other direction, suppose $\W'_P(I) \subseteq I$.
Let $I^- = \{ a ~|~ not~a \in I \}$ be the atoms that evaluate to $\false$ in $I$.
Then $\Phi^+_P(I) \subseteq I$, $\Phi^-_P(I) \subseteq I^-$, and $\Z_{P}(I) \subseteq I^-$.
Let $S = \U_{P}(I) \backslash I^-$.
For every $a \in S$, $ a \notin \Phi^-_P(I)$ because $ \Phi^-_P(I) \subseteq I^-$, so
there is a rule in $ground(P)$ such that no body literal evaluates to $\false$ under $I$
and, hence, there is a body atom $b$ in $\U_{P}(I)$,
by the definition of unfounded set.
Further, $b \notin I^{-}$, because no body literal evaluates to $\false$ under $I$.
Thus, $b \in S$.
Hence, $S$ is a circular unfounded set of $P$ with respect to $I$.
Thus, $S \subseteq \Z_{P}(I) \subseteq I^-$.
This shows that $S$ must be empty.
Hence, $\U_{P}(I) \subseteq I^-$ and, consequently,
$\W_P(I) \subseteq I$.

\noindent
Part 2.
Suppose $\W'_P(I) = I$. 
Then $\W_P(I) \subseteq I$ by Part 1.
$\W'_P(I) \subseteq \W_P(I)$, as observed above, so $I \subseteq \W_P(I)$.
Hence, $\W_P(I) = I$.

Suppose $\W_P(I) = I$. 
Then $\W'_P(I) \subseteq I$ by Part 1.
Let $S = \U_P(I) \backslash \Phi^-_P(I)$.
Suppose $a \in S$.
Then there is a rule instance $a \mbox{ :- } B$
such that $I(B) \neq \false$, because $a \notin \Phi^-_P(I)$,
and, hence, there is an atom $b \in B$ such that $b \in \U_P(I)$, by the definition of unfounded set.
But then $not~b \in I$ because $I = \W_P(I) = \Phi^+_P(I) \cup \U_P(I)$
and, as a result, $I(B) = \false$.
This contradiction shows that $S$ is empty.
Consequently, $\U_P(I) = \Phi^-_P(I)$.

Now $I = \W_P(I) = \Phi^+_P(I) \cup \U_P(I) = \Phi^+_P(I) \cup \Phi^-_P(I)  \subseteq \W'_P(I)$.
Hence, $\W'_P(I) = I$.
\end{proof}
}

It is now straightforward to show that $\W'_P$ provides an alternative formulation of the
well-founded semantics.
\begin{theorem}   \label{thm:equiv}
For any interpretation $J$, if either $lfp(\W_P, J)$ or $lfp(\W'_P, J)$ is defined, then they are both defined and
\[
lfp(\W_P, J) = lfp(\W'_P, J)
\]
In particular
\[
lfp(\W_P) = lfp(\W'_P)
\]
\end{theorem}
\skipit{
\begin{proof}
If there is a least fixedpoint greater than $J$,
then it is a fixedpoint of both $\W_P$ and $\W'_P$, by the previous lemma.
In particular, when $J = \emptyset$, $lfp(\W_P) = lfp(\W'_P)$.
\end{proof}
}

Because $\Phi_P^+$, $\Phi_P^-$, and $\Z_P$ are monotonic,
they can be interleaved in any fair way to obtain the fixedpoint,
as a chaotic iteration.
Indeed, each of these functions can be decomposed further,
and application of those pieces interleaved.

Finally, using the alternate characterization of the well-founded semantics,
we have a simple proof of an extension of a result of \cite{Berman95}.
If a logic program is positive order-consistent (or tight) then the well-founded and Fitting semantics coincide.

\begin{propn}    \label{prop:tight}
Let $P$ be a logic program, 
$\P \subseteq \Pi$ be a downward-closed set of predicates with floor $\F$,
$\Q$ be  $\P\backslash\F$, and
$Q$ be the rules in $P$ defining predicates in $\Q$.
Let $I$ be a fixed semantics for $\F$.

If $Q$ is positive order-consistent, then 
$lfp(\W_Q, I) = lfp(\Phi_Q, I)$.

\end{propn}
\skipit{
\begin{proof}
$Q$ is positive order-consistent, so $\leq_0$ is well-founded
and $\Z_{Q}(J)$ is empty, for every $J$.
Consequently $\W'_Q(J) = \Phi_Q(J)$, for every $J$, and, hence,
$lfp(\W_Q, I) = lfp(\Phi_Q, I)$.
\end{proof}
}

\section{Results}   \label{sec:results}

If $\P$ is downward-closed with floor $\F$,
and $I$ is a 3-valued Herbrand model of predicates in $\F$
then  $lfp( \W_P, I)$ is the well-founded model of $P$ extending $I$, 
that is, the well-founded semantics of $P$ after all predicates in $\F$ are interpreted according to $I$.
Similarly,
$lfp( \Phi_P, I)$, the Fitting semantics of $P$ extending $I$,
is the Fitting semantics of $P$ after all predicates in $\F$ are interpreted according to $I$.
Such a notion is interesting, in general, because predicates in $\F$ might be defined outside
the logic programming setting, or in a different module.

It is interesting in the context of this paper because often a full logic program does not have a signing
but, after omitting some support and utility predicates, the remainder does have a signing.
So, if $\P$ is downward-closed with floor $\F$, then $\F$ can be considered the support predicates
and $\P\backslash\F$ the remainder with a signing.

We can now establish conditions under which the well-founded semantics and Fitting semantics agree
on the truth value of some ground literals
(even if they may disagree on other literals).
This can be seen as a refinement of Proposition~\ref{prop:tight} in that only a subset of predicates
is required to be positive order-consistent.

\begin{theorem}    \label{thm:EQ}
Let $P$ be a logic program, 
$\P \subseteq \Pi$ be a downward-closed set of predicates with floor $\F$,
$\Q$ be  $\P\backslash\F$, and
$s$ be a signing on $\Q$.
Let $I$ be a fixed semantics for $\F$ and  $p$ be a predicate defined in $\Q$.
Suppose  $p$ avoids negative unfoundedness wrt $s$.

\noindent
For any ground atom $p(\vec{a})$:

If $s(p) = +1$ then
\phantom{not~} $p(\vec{a}) \in lfp( \W_Q, I)$  ~  iff    \, \phantom{not~}    $p(\vec{a}) \in lfp( \Phi_Q, I)$ 

If $s(p) = -1$ then
$not~ p(\vec{a}) \in lfp( \W_Q, I)$   ~  iff   ~  $not~ p(\vec{a}) \in lfp( \Phi_Q, I)$ 

\end{theorem}
\skipit{
\begin{proof}

Let $Q$ denote the subset of $P$ defining predicates $q$ of $\Q$ that $p$ depends on (that is, $p \geq q$).
It suffices to prove the result for $Q$, 
since $P$ and $Q$ agree on the semantics of the predicates in $\Q$
under both well-founded and Fitting semantics.

We  prove the theorem using transfinite induction on the Kleene sequences from $I$.
The ``if'' directions are well-known to hold when the floor is empty, independent of the other conditions \cite{WF91}.
They can be proved by transfinite induction, using the facts that
$\Phi_P(I ) \subseteq \W_P(I)$, for any $P$ and $I$, and both functions are monotonic over 
the subset ordering.

The proof of the ``only if'' directions is also by transfinite induction on the Kleene sequence.
We use the alternate formulation of the well-founded semantics.
We claim that, at each stage $\alpha$, 
the set of true positively-signed atoms is the same in $\Phi_Q \uparrow \alpha$ and $\W'_Q \uparrow \alpha$
and the set of false negatively-signed atoms is the same in $\Phi_Q \uparrow \alpha$ and $\W'_Q \uparrow \alpha$.
This is true trivially when $\alpha = 0$, since $\Phi_Q \uparrow 0 = I = \W'_Q \uparrow 0$.
If $\alpha$ is a limit ordinal then $\Phi_Q \uparrow \alpha = \bigcup_{\beta < \alpha} \Phi_Q \uparrow \beta$
and $\W'_Q \uparrow \alpha = \bigcup_{\beta < \alpha} \W'_Q \uparrow \beta$.
By the induction hypothesis, the two sets agree on true positively-signed atoms and false negatively-signed atoms.

Suppose $\alpha$ is a successor ordinal, say $\alpha = \beta + 1$.
Suppose the induction hypothesis holds at $\beta$.
Consider a positively-signed predicate $p$ and a ground instance
$a \mbox{ :- } b_1, \ldots, b_n, not~c_1, \ldots, not~c_m$ of a rule for $p$.
Because of the signing, $b_1, \ldots, b_n$ have positive signs and $c_1, \ldots, c_m$ have negative signs,
except perhaps for those with predicates from $\F$.
By the induction hypothesis, $b_i \in \W'_Q \uparrow \beta$ iff $b_i \in \Phi_Q \uparrow \beta$ and
$not~ c_j \in \W'_Q \uparrow \beta$ iff $not~  c_j \in \Phi_Q \uparrow \beta$.
Hence, $a \in \W'_Q \uparrow \alpha$ iff $a \in \Phi_Q \uparrow \alpha$.

Now, consider a negatively-signed predicate $p$ and a ground instance
of a rule for $p$: 
\linebreak
$a \mbox{ :- } b_1, \ldots, b_n, not~ c_1, \ldots, not~c_m$.
Under the signing $s$, $b_1, \ldots, b_n$ have negative signs and $c_1, \ldots, c_m$ have positive signs,
except for those with predicates from $\F$.
Because $p$ avoids negative unfoundedness wrt $s$,
there is not a circular unfounded set wrt $\W'_Q \uparrow \beta$ containing any $b_i$.
Thus, $not~ a \in  \W'_Q(\W'_Q \uparrow \beta)$ iff $not~ a \in \neg \Phi^-_Q(\W'_Q \uparrow \beta)$ iff 
some $not~ b_i \in \W'_Q \uparrow \beta$ or
some $c_j \in \W'_Q \uparrow \beta$, for every ground instance of a rule in $P$ with head $a$.
By the induction hypothesis, $not~ b_i \in \W'_Q \uparrow \beta$ iff $not~ b_i \in \Phi_Q \uparrow \beta$ and
$c_j \in \W'_Q \uparrow \beta$ iff $c_j \in \Phi_Q \uparrow \beta$.
Hence, $not~a \in \W'_Q \uparrow (\beta+1)$ iff $not~a \in \Phi_Q \uparrow (\beta+1)$.

By induction, the claim holds for each stage $\alpha$ and, in particular, at the fixedpoint.
Thus, we have
$p(\vec{a}) \in lfp( \W_Q, I)$   iff    $p(\vec{a}) \in lfp( \Phi_Q, I)$ 
for any predicate $p$ in $\P$ with a positive sign,
and also
$not~ p(\vec{a}) \in lfp( \W_Q, I)$   iff    $not~ p(\vec{a}) \in lfp( \Phi_Q, I)$ 
for any predicate $p$ in $\P$ with a negative sign.
\end{proof}
}

Hence, if we are only interested in the part of a predicate $p$ satisfying the conditions of the theorem,
and the predicates of $\F$ are computed, then
the Fitting and well-founded semantics are equivalent.
That is, it suffices to compute using $\Phi_P$ instead of $\W_P$ or $\W'_P$.
When $P$ is a finite propositional program, this gives a linear cost to compute the part of $p$,
instead of a worst-case quadratic cost.
(In other cases, the advantage is less clear.)
Alternatively, under the conditions of the theorem,
we can compute Fitting's semantics for the part of $p$ by using an implementation of the well-founded semantics.

\begin{figure}[t]

\begin{tikzpicture}[->,>=stealth',shorten >=1pt,auto,node distance=2.6cm,semithick]

  \tikzstyle{every state}=[fill=none,draw=none,text=black]

  \node[state] (Dbly)							{\texttt{defeasibly}};
  \node[state] (Over)	[below right of=Dbly]			{\texttt{overruled}};
  \node[state] (Dftd)	[below right of=Over]			{\texttt{defeated}};
  \node[state] (Lbda)	[below  left of=Over]			{\texttt{lambda}};
  \node[state] (Dfly)	[below left of=Lbda]			{\texttt{definitely}};
  \node[state] (Sup)	[below right of=Dftd]			{\texttt{sup}};
  \node[state] (Fact)	[below  left of=Dfly]			{\texttt{fact}};
  \node[state] (Rule)	[below       of=Dftd]			{\texttt{rule}};
  \node[state] (SorD)	[below left of=Rule]			{\texttt{s\_or\_d}};	
  \node[state] (Dftr)	[below right of=Rule]			{\texttt{defeater}};
  \node[state] (Def)	[below right of=SorD]		{\texttt{defeasible}};
  \node[state] (Str)	[below left of=SorD]			{\texttt{strict}};

  \path (Dbly)	edge 				node {\textbf{--}}	(Over)
  	 		edge 				node {+\textbf{--}}	(Dfly)
 			edge 	[loop above]	node {+}	(Dbly)
   	(Over)  	edge 				node {+}	(Lbda)
   			edge 				node {\textbf{--}}	(Dftd)
   			edge 				node {+}	(Rule)
   	(Lbda)  	edge 				node {+\textbf{--}} (Dfly)
   			edge 				node {+}	(SorD)
			edge 	[loop right]		node {+}	(Lbda)
   	(Dftd)  	edge 				node {+}	(Sup)
			edge 				node {+}	(SorD)
   			edge 	[bend right]	node {+}	(Dbly)
   (Rule)  	edge 				node {+}	(SorD)
   		edge 				node {+}	(Dftr)
   (SorD)  	edge 				node {+}	(Str)
   		edge 				node {+}	(Def)
   (Dfly)  	edge 				node {+}	(Str)
   		edge 				node {+}	(Fact)
  		edge 	[loop left]		node {+}	(Dfly);

\end{tikzpicture}
\caption{
\label{fig:pdg}
A signed predicate dependency graph.  Positive (negative) dependencies are labelled by + (respectively, --).}
\end{figure}
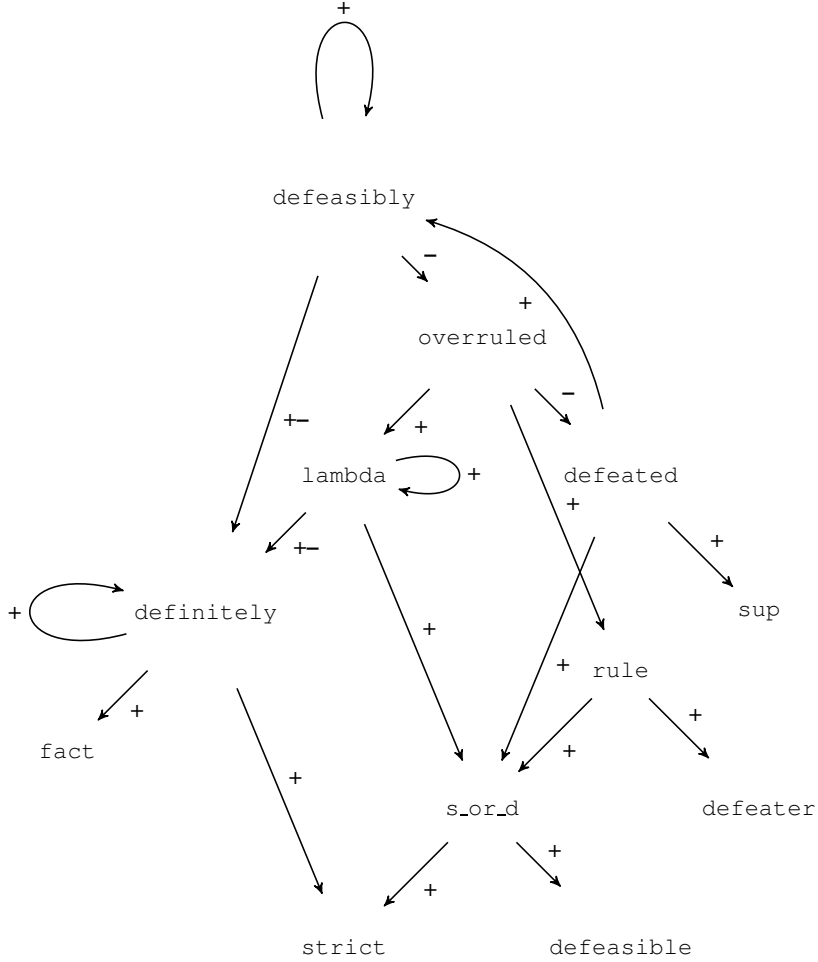

This result is applicable to the metaprogram for the defeasible logic $\DL(\pl)$ \cite{sdl2},
where we are only interested in the positive part of the predicate $\mt{defeasibly}$.
That part represents the defeasible consequences of a defeasible theory.
The signed program dependency graph is depicted in Figure \ref{fig:pdg}.
Let $\P$ be all predicates in the program 
and $\Q$ be the set $\{ \mt{defeasibly},$ $ \mt{overruled}, \mt{defeated} \}$.
Let $\F$ be $\P \backslash \Q$.
Then $\P$ is a downward-closed set of predicates with floor $\F$.
Furthermore, $\Q$ has a signing $s$ that maps
$\mt{defeasibly}$ and $\mt{defeated}$  to $+1$
and $\mt{overruled}$ to $-1$.
Finally, 
$\mt{overruled}$ avoids negative unfoundedness wrt $s$ and $\Q$,
because $\mt{overruled}$ does not depend positively on any predicate in $\Q$.

The previous theorem shows that, in some circumstances,
we can ignore the computation of unfounded sets when computing the well-founded semantics.
In the next theorem we show that, in different circumstances,
only the positive or only the negative conclusions for each predicate need to be computed.
In terms of an alternating fixedpoint implementation \cite{altfpt},
only the underestimate or only the overestimate needs to be computed for each predicate.

We will need refinements of the $\Phi$ functions used in defining Fitting's semantics.
Let $s$ be a signing on $\P$.
We define $\Phi^{+s}_P(I) = \{ a \in \Phi^+_P(I) ~|~ s(a) = +1 \}$
and $\Phi^{-s}_P(I) = \{ a \in \Phi^-_P(I) ~|~ s(a) = -1 \}$.
Notice that $+$ and $-$ in this notation are doing double duty,
representing both the sign of the predicates and the polarity of the inference.

Now we can define a very weak form of $\W'_P$:
\[
\UU_P^s(I) = \Phi^{+s}_P(I) \cup \neg \Phi^{-s}_P(I) \cup \neg \Zm_P(I)
\]
$\UU_P^s$ only infers positively signed atoms and negatively signed negative literals,
that is, it only infers the positive part of a positively signed predicate
and the negative part of a negatively signed predicate.
So $\UU_P^s$ is substantially weaker than $\W'_P$.
In fact, $\W'_P(I) = \UU_P^s(I) \cup \UU_P^{\bar{s}}(I)$.

\begin{theorem}    \label{thm:new}
Let $P$ be a logic program, 
$\P \subseteq \Pi$ be a downward-closed set of predicates with floor $\F$,
$\Q$ be  $\P\backslash\F$, and
$s$ be a signing on $\Q$.
Let $I$ be a fixed semantics for $\F$.

\noindent
For any ground atom $a$:

If $s(a) = +1$ then
\phantom{not~} $a \in lfp(\W_Q, I)$ ~ iff ~\phantom{not~}  $a \in lfp(\UU_Q^s, I)$

If $s(a) = -1$ then
$not~a \in lfp(\W_Q, I)$ ~ iff ~ $not~a \in lfp(\UU_Q^s, I)$

\end{theorem}
\skipit{
\begin{proof}
Let $Q$ be the clauses defining the predicates in $\Q$.
We use the alternate formulation of the well-founded semantics.
We claim that, at each stage $\alpha$, 
the set of true positively-signed atoms is the same in $\UU_Q^s \uparrow \alpha$ and $\W'_Q \uparrow \alpha$
and the set of false negatively-signed atoms is the same in $\UU_Q^s \uparrow \alpha$ and $\W'_Q \uparrow \alpha$.
This is true trivially when $\alpha = 0$, since $\UU_Q^s \uparrow 0 = I = \W'_Q \uparrow 0$.
If $\alpha$ is a limit ordinal then $\UU_Q^s \uparrow \alpha = \bigcup_{\beta < \alpha} \UU_Q^s \uparrow \beta$
and $\W'_Q \uparrow \alpha = \bigcup_{\beta < \alpha} \W'_Q \uparrow \beta$.
By the induction hypothesis, the two sets agree on true positively-signed atoms and false negatively-signed atoms.

If $\alpha$ is a successor ordinal, say $\alpha = \beta + 1$,
suppose the induction hypothesis holds at $\beta$.
Consider any positively-signed predicate $p$ and a ground instance
$a \mbox{ :- } b_1, \ldots, b_n, not~c_1, \ldots, not~c_m$ of a rule for $p$.
Because of the signing, $b_1, \ldots, b_n$ have positive signs and $c_1, \ldots, c_m$ have negative signs,
except perhaps for those predicates from $\F$, which are not signed.
By the induction hypothesis, $b_i \in \W'_Q \uparrow \beta$ iff $b_i \in \UU^s_Q \uparrow \beta$ and
$not~ c_j \in \W'_Q \uparrow \beta$ iff $not~  c_j \in \UU^s_Q \uparrow \beta$.
Hence, $a \in \W'_Q \uparrow (\beta + 1)$ iff $a \in \UU^s_Q \uparrow (\beta + 1)$.

Now, consider a negatively-signed predicate $p$ and a ground instance
of a rule for $p$:
\linebreak
$a \mbox{ :- } b_1, \ldots, b_n, not~ c_1, \ldots, not~c_m$. 
Under the signing $s$, $b_1, \ldots, b_n$ have negative signs and 
\linebreak
$c_1, \ldots, c_m$ have positive signs,
again except for predicates in $\F$.
By the induction hypothesis, 
$not~b_i  \in \W'_Q \uparrow \beta$ iff $not~b_i \in \UU^s_Q \uparrow \beta$ and
$c_j \in \W'_Q \uparrow \beta$ iff $c_j \in \UU^s_Q \uparrow \beta$.
Since this applies to all such clauses for $a$,  
$a \in \Phi^{+}_P( \W'_Q \uparrow \beta )$ iff $a \in \Phi^{+s}_P(  \UU^s_Q \uparrow \beta )$.
Furthermore,
$a \in \Z_Q ( \W'_Q \uparrow \beta )$ iff $a \in \Zm_Q (  \UU^s_Q \uparrow \beta )$,
because $a$ has a negative sign.
Hence, $a \in \W'_Q \uparrow (\beta + 1)$ iff $a \in \UU^s_Q \uparrow (\beta + 1)$.
This applies to all atoms $a$ with a negatively-signed predicate.

By transfinite induction, for all ordinals $\alpha$, and all ground atoms $a$,
if $s(a) = +1$ then
$a \in \W'_Q \uparrow \alpha$ iff $a \in \UU_Q^s \uparrow \alpha$,
and
if $s(a) = -1$ then
$not~a \in \W'_Q \uparrow \alpha$  iff  $not~a \in \UU_Q^s \uparrow \alpha$.
In particular, this applies to the ordinal that reaches a fixedpoint.

Since $lfp(\W_Q, I) = lfp(\W'_Q, I)$ (by Theorem \ref{thm:equiv}), the result follows.
\end{proof}
}

Thus, for signed programs, the situation is analogous to computing predicates in stratified programs,
where only the positive part need be computed: 
the negative part is then known as the complement of the positive part.
However, in the case of signed programs, the negative part is not needed.

If we want to infer a negatively signed atom, we can use  $\bar{s}$ in place of $s$,
that is, use $\UU_P^{\bar{s}}$.
However, if we want to infer both a negatively signed atom and a positively signed atom,
even if they contain different predicates, neither function will suffice.

The above theorem suggests a very simple analysis for arbitrary programs that, given a query,
can determine whether the positive or negative parts of a predicate  do not need to be computed
when evaluating the well-founded semantics.
Specifically, we identify which predicates appear positively and which appear negatively in the query,
that is, which predicates require their positive part and which require their negative part.
This information can then be propagated over the signed predicate dependency graph.
If the positive part of $p$ may be needed and $p$ depends positively on $q$,
then the positive part of $q$ may be needed.
If the positive part of $p$ may be needed and $p$ depends negatively on $q$,
then the negative part of $q$ may be needed.
Similarly, if the negative part of $p$ may be needed and $p$ depends positively (negatively) on $q$,
then the negative (positive) part of $q$ may be needed.

After full propagation, those predicates that may need only the positive (respectively, negative) parts
to be computed can safely avoid generating their negative (respectively, positive) parts.
Those predicates that may be needed both positively and negatively will need both parts to be computed.
This is a simple analysis, and should be easy to incorporate into an algorithm for answering queries under the well-founded semantics.
Its cost is linear in the size of the signed predicate dependency graph.
For implementations based on Van Gelder's alternating fixedpoint approach,
the result should be an improvement in both time and space requirements.

Applying this analysis to the program of Figure \ref{fig:pdg},
where only the positive part of $\mt{defeasibly}$ is of interest,
we find that only the predicates 
$\mt{definitely}$, $\mt{fact}$, $\mt{s\_or\_d}$, $\mt{defeasible}$, and $\mt{strict}$
may be needed both positively and negatively.
For each of the remaining predicates, only one part will need to be computed.

Specifically, $\mt{definitely}$ and $\mt{lambda}$ can be computed using only their positive parts,
since that part of the program is stratified.  Then we only need the positive part of $\mt{defeated}$
and the negative part of $\mt{overruled}$ to compute the positive part of $\mt{defeasibly}$.


The integration of the ideas in this paper with magic sets \cite{Morishita,jELS}
and optimizations based on pre-mappings \cite{prem} 
is left for further research.





\textbf{Acknowledgements:}
The author has an adjunct position at Griffith University and an honorary position at UNSW.
He thanks the referees for comments that helped improve this paper.

\bibliography{signing}

\begin{thebibliography}{}

\bibitem[\protect\citeauthoryear{Apt, Blair, and Walker}{Apt
  et~al\mbox{.}}{1988}]{ABW}
{\sc Apt, K.~R.}, {\sc Blair, H.~A.}, {\sc and} {\sc Walker, A.} 1988.
\newblock Towards a theory of declarative knowledge.
\newblock In {\em Foundations of Deductive Databases and Logic Programming.}
  Morgan Kaufmann, 89--148.

\bibitem[\protect\citeauthoryear{Apt and Bol}{Apt and Bol}{1994}]{AptBol}
{\sc Apt, K.~R.} {\sc and} {\sc Bol, R.~N.} 1994.
\newblock Logic programming and negation: A survey.
\newblock {\em J. Log. Program.\/}~{\em 19/20}, 9--71.

\bibitem[\protect\citeauthoryear{Berman, Schlipf, and Franco}{Berman
  et~al\mbox{.}}{1995}]{Berman95}
{\sc Berman, K.~A.}, {\sc Schlipf, J.~S.}, {\sc and} {\sc Franco, J.~V.} 1995.
\newblock Computing well-founded semantics faster.
\newblock In {\em Logic Programming and Nonmonotonic Reasoning, Third
  International Conference, LPNMR'95}, {V.~W. Marek} {and} {A.~Nerode}, Eds.
  Lecture Notes in Computer Science, vol. 928. Springer, 113--126.

\bibitem[\protect\citeauthoryear{Brass, Dix, Freitag, and Zukowski}{Brass
  et~al\mbox{.}}{2001}]{trans_BU}
{\sc Brass, S.}, {\sc Dix, J.}, {\sc Freitag, B.}, {\sc and} {\sc Zukowski, U.}
  2001.
\newblock Transformation-based bottom-up computation of the well-founded model.
\newblock {\em Theory and Practice of Logic Programming\/}~{\em 1,\/}~5,
  497–538.

\bibitem[\protect\citeauthoryear{Cormen, Leiserson, Rivest, and Stein}{Cormen
  et~al\mbox{.}}{2001}]{Cormen}
{\sc Cormen, T.~H.}, {\sc Leiserson, C.~E.}, {\sc Rivest, R.~L.}, {\sc and}
  {\sc Stein, C.} 2001.
\newblock {\em Introduction to Algorithms, Second Edition}.
\newblock The {MIT} Press and McGraw-Hill Book Company.

\bibitem[\protect\citeauthoryear{Drabent and Martelli}{Drabent and
  Martelli}{1991}]{strictcomp}
{\sc Drabent, W.} {\sc and} {\sc Martelli, M.} 1991.
\newblock Strict completion of logic programs.
\newblock {\em New Gener. Comput.\/}~{\em 9,\/}~1, 69--80.

\bibitem[\protect\citeauthoryear{Dung}{Dung}{1992}]{Dung92}
{\sc Dung, P.~M.} 1992.
\newblock On the relations between stable and well-founded semantics of logic
  programs.
\newblock {\em Theor. Comput. Sci.\/}~{\em 105,\/}~1, 7--25.

\bibitem[\protect\citeauthoryear{Fages}{Fages}{1994}]{Fages94}
{\sc Fages, F.} 1994.
\newblock Consistency of {Clark's} completion and existence of stable models.
\newblock {\em Meth. of Logic in {CS}\/}~{\em 1,\/}~1, 51--60.

\bibitem[\protect\citeauthoryear{Fitting}{Fitting}{1985}]{Fitting}
{\sc Fitting, M.} 1985.
\newblock A {K}ripke-{K}leene semantics for logic programs.
\newblock {\em J. Log. Program.\/}~{\em 2,\/}~4, 295--312.

\bibitem[\protect\citeauthoryear{Gelfond and Lifschitz}{Gelfond and
  Lifschitz}{1988}]{stable}
{\sc Gelfond, M.} {\sc and} {\sc Lifschitz, V.} 1988.
\newblock The stable model semantics for logic programming.
\newblock In {\em Proc. JICSLP}. 1070--1080.

\bibitem[\protect\citeauthoryear{Gire}{Gire}{1992}]{Gire92}
{\sc Gire, F.} 1992.
\newblock Well founded semantics and stable semantics of semi-strict programs.
\newblock In {\em Database Theory - ICDT'92, 4th International Conference,
  Berlin, Germany, October 14-16, 1992, Proceedings}, {J.~Biskup} {and}
  {R.~Hull}, Eds. Lecture Notes in Computer Science, vol. 646. Springer,
  261--275.

\bibitem[\protect\citeauthoryear{Gire}{Gire}{1994}]{Gire94}
{\sc Gire, F.} 1994.
\newblock Equivalence of well-founded and stable semantics.
\newblock {\em J. Log. Program.\/}~{\em 21,\/}~2, 95--111.

\bibitem[\protect\citeauthoryear{Hitzler and Wendt}{Hitzler and
  Wendt}{2002}]{HW02}
{\sc Hitzler, P.} {\sc and} {\sc Wendt, M.} 2002.
\newblock The well-founded semantics is a stratified {Fitting} semantics.
\newblock In {\em {KI} 2002: Advances in Artificial Intelligence, 25th Annual
  German Conference on AI, {KI} 2002}, {M.~Jarke}, {J.~Koehler}, {and}
  {G.~Lakemeyer}, Eds. Lecture Notes in Computer Science, vol. 2479. Springer,
  205--221.

\bibitem[\protect\citeauthoryear{Kemp, Ramamohanarao, and Stuckey}{Kemp
  et~al\mbox{.}}{1997}]{jELS}
{\sc Kemp, D.~B.}, {\sc Ramamohanarao, K.}, {\sc and} {\sc Stuckey, P.~J.}
  1997.
\newblock An efficient evaluation technique for non-stratified programs by
  transformation to explicitly locally stratified programs.
\newblock {\em Journal of Systems Integration\/}~{\em 7,\/}~3/4, 191--230.

\bibitem[\protect\citeauthoryear{Kleene}{Kleene}{1952}]{Kleene}
{\sc Kleene, S.~C.} 1952.
\newblock {\em Introduction to Metamathematics}.
\newblock Van Nostrand.

\bibitem[\protect\citeauthoryear{Kunen}{Kunen}{1989}]{Kunen89}
{\sc Kunen, K.} 1989.
\newblock Signed data dependencies in logic programs.
\newblock {\em J. Log. Program.\/}~{\em 7,\/}~3, 231--245.

\bibitem[\protect\citeauthoryear{Maher}{Maher}{2021}]{sdl2}
{\sc Maher, M.~J.} 2021.
\newblock Defeasible reasoning via {Datalog$^\neg$}.
\newblock {\em {}\/}, forthcoming.

\bibitem[\protect\citeauthoryear{Maher, Tachmazidis, Antoniou, Wade, and
  Cheng}{Maher et~al\mbox{.}}{2020}]{sdl}
{\sc Maher, M.~J.}, {\sc Tachmazidis, I.}, {\sc Antoniou, G.}, {\sc Wade, S.},
  {\sc and} {\sc Cheng, L.} 2020.
\newblock Rethinking defeasible reasoning: A scalable approach.
\newblock {\em {TPLP}\/}~{\em 20,\/}~4, 552--586.

\bibitem[\protect\citeauthoryear{Morishita}{Morishita}{1996}]{Morishita}
{\sc Morishita, S.} 1996.
\newblock An extension of {Van Gelder's} alternating fixpoint to magic
  programs.
\newblock {\em J. Comput. Syst. Sci.\/}~{\em 52,\/}~3, 506--521.

\bibitem[\protect\citeauthoryear{Turner}{Turner}{1993}]{Turner93}
{\sc Turner, H.} 1993.
\newblock A monotonicity theorem for extended logic programs.
\newblock In {\em Proc. of the Tenth Int. Conf. on Logic Programming}.
  567--585.

\bibitem[\protect\citeauthoryear{{Van~Gelder}}{{Van~Gelder}}{1989}]{altfpt}
{\sc {Van~Gelder}, A.} 1989.
\newblock The alternating fixpoint of logic programs with negation.
\newblock In {\em Proceedings of the Eighth ACM Symposium on Principles of
  Database Systems}. 1--10.

\bibitem[\protect\citeauthoryear{{Van~Gelder}, Ross, and Schlipf}{{Van~Gelder}
  et~al\mbox{.}}{1991}]{WF91}
{\sc {Van~Gelder}, A.}, {\sc Ross, K.~A.}, {\sc and} {\sc Schlipf, J.~S.} 1991.
\newblock The well-founded semantics for general logic programs.
\newblock {\em J. ACM\/}~{\em 38,\/}~3, 620--650.

\bibitem[\protect\citeauthoryear{Zaniolo, Yang, Interlandi, Das, Shkapsky, and
  Condie}{Zaniolo et~al\mbox{.}}{2017}]{prem}
{\sc Zaniolo, C.}, {\sc Yang, M.}, {\sc Interlandi, M.}, {\sc Das, A.}, {\sc
  Shkapsky, A.}, {\sc and} {\sc Condie, T.} 2017.
\newblock Fixpoint semantics and optimization of recursive {Datalog} programs
  with aggregates.
\newblock {\em {TPLP}\/}~{\em 17,\/}~5-6, 1048--1065.

\end{thebibliography}
\vfill

\appendix

\pagebreak

\section{Re-proofs}
This appendix contains proofs of minor variations of theorems in \cite{Kunen89}, \cite{Dung92}, and \cite{Turner93}.
This is just a bonus; the results of this paper do not depend on these results.
The statements of the theorems now allow a floor, which is not signed, 
and have other small variations.

The following theorem is a minor variation of a theorem (Theorem 3.3)
that was stated in \cite{Kunen89} without proof.

\removebrackets
\begin{theorem}[\cite{Kunen89}]    \label{thm:Kunen}
Let $P$ be a logic program, 
$\P \subseteq \Pi$ be a downward-closed set of predicates with floor $\F$,
let $\Q$ be  $\P\backslash\F$, and
$s$ be a signing on $\Q$ for $P$.
Let $I$ be a fixed two-valued semantics for $\F$ and  
let $A$ be a 3-valued model of Clark's completion $P^*$ extending $I$.
Then there is a two-valued model $A'$ of $P^*$ that extends $A$.

\noindent
Specifically, $A'$ can be defined as follows:

If $s(p) = +1$ then
$\phantom{not~} p(\vec{a}) \in A'$  iff $not~p(\vec{a}) \notin A$

If $s(p) = -1$ then
$not~p(\vec{a}) \in A'$ iff $\phantom{not~} p(\vec{a}) \notin A$
\end{theorem}
\skipit{
\begin{proof}
We need to establish that $A'$ is a model of $P^*$, Clark's completion of $P$.
Consider $ground(P)$, where the rules of $P$ are grounded by the elements of $A$,
and the corresponding completion $ground(P)^*$, which might involve infinite disjunctions.
We will refer to the elements of $ground(P)^*$ of the form $a \leftrightarrow \vee_{i \in I} B_i$ as \emph{defs}.

For each ground atom $a$, consider all ground rules with $a$ as head, and the corresponding def.
If $A(a) = \true$ then the body of some rule evaluates to $\true$ in $A$.
If $A(a) = \false$ then in each rule, some literal evaluates to $\false$ in $A$.
In both cases, the def for $a$ is also satisfied by $A'$.

If $A(a) = \unknown$ then no rule body evaluates to $\true$ and some rule body $B$ evaluates to $\unknown$
(that is, no body literal evaluates to $\false$ and at least one literal evaluates to $\unknown$) in $A$.
Suppose $s(a) = +1$.  Then $A'(a) = \true$, from the definition of $A'$.  
If an atom $b \in B$ evaluates to $\unknown$ in $A$ then, because $s(b) = +1$, $A'(b) = \true$.
If $not~c \in B$  evaluates to $\unknown$ in $A$  then, because $s(c) = -1$, $A'(c) = \false$.
As a result, $A'(B) = \true$.
Now suppose $s(a) = -1$.  Then $A'(a) = \false$.  
If an atom $b \in B$ evaluates to $\unknown$ in $A$ then, because $s(b) = -1$, $A'(b) = \false$.
If $not~c \in B$  evaluates to $\unknown$ in $A$  then, because $s(c) = +1$, $A'(c) = \true$.
As a result, $A'(B) = \false$.
In both cases, the def for $a$ is satisfied by $A'$.
\end{proof}
}

When $I$ is not already a 2-valued model of $P^*$, we can get a second 2-valued model
by using $\bar{s}$, instead of $s$.
This is similar to Turner's extension of a well-founded model to a stable model, below,
where $\bar{s}$ provides a second stable model, provided the well-founded model is not stable itself.


To present Turner's theorem, we need to define stable models.
Let $P$ be a ground logic program and $I$ be a 2-valued interpretation.
Then the Gelfond-Lifschitz reduct of $P$ wrt $I$, denoted by $P^I$,
is obtained by deleting from $P$ those rules with a negative literal that evaluates to $\false$ in $I$,
and deleting those negative literals that evaluate to $\true$ in $I$ from the remaining rules.
A stable model is a 2-valued interpretation $S$ such that the least model of $P^S$ is $S$ \cite{stable}.

The proof of Theorem 2 of \cite{Turner93} is very compact and in a notation I am not familiar with.
The following proof of a variation of that theorem is also brief, and is more intuitive to me.

\removebrackets
\begin{theorem}[\cite{Turner93}]    \label{thm:Turner}
Let $P$ be a logic program, 
$\P \subseteq \Pi$ be a downward-closed set of predicates with floor $\F$,
let $\Q$ be  $\P\backslash\F$, and
$s$ be a signing on $\Q$ for $P$.
Let $I$ be a fixed stable model for $\F$ and  
let $W$ be the well-founded model of $P$ extending $I$.
Then there is a stable model $S$ of $P$ that extends $W$.

\noindent					
Specifically, $S$ can be defined as follows:

If $s(p) = +1$ then
$\phantom{not~} p(\vec{a}) \in S$  iff $not~p(\vec{a}) \notin W$

If $s(p) = -1$ then
$not~p(\vec{a}) \in S$ iff $\phantom{not~} p(\vec{a}) \notin W$
\end{theorem}
\skipit{
\begin{proof}
We need to prove that $S$ is a stable model.
$W$ is a 3-valued model of $P^*$ \cite{WF91} so,
by Theorem \ref{thm:Kunen}, $S$ is a 2-valued model of $P^*$ and, hence, also a model of $P$.

We use a characterization of stable models established in Theorem 2.5 of \cite{Dung92}. 
A model $S$ of $P$ is stable iff for all $A$,  
if $A$ is unfounded wrt $S$ then $A \cap S = \emptyset$.

We first prove by induction on the Kleene sequence that no atom $a \in W$ is in an unfounded set wrt $S$.
If $a \in \W \uparrow (\beta + 1)$ then there is a rule $a \mbox{ :- } B$ in $ground(P)$ such that 
$B$ evaluates to $\true$ in $\W \uparrow \beta$.  
By the induction hypothesis, every atom $b \in B$ is not in an unfounded set wrt $S$.
Hence, $a$ is not in an unfounded set wrt $S$,
since $B$ is neither $\false$ in $S$ nor does $B$ contain an atom from an unfounded set wrt $S$.
If $a \in \W \uparrow \alpha$, for limit ordinal $\alpha$, then $a \in \W \uparrow \beta$ for some $\beta < \alpha$ and,
by the induction hypothesis, $a$ is not in an unfounded set wrt $S$.
Hence, if $A$ is unfounded wrt $S$, then $A \subseteq S \backslash W$. 

Let $A$ be unfounded wrt $S$.
Then, for any $a \in A$, $a \in S \backslash W$ and, by the definition of $S$, $s(a) = +1$.
Furthermore, for any rule $a \mbox{ :- } B$ for $a$, if $b \in B$ then $s(b)=+1$ and if $not~b \in B$ then $s(b)=-1$.
Thus, for any $a \in A$ and rule $a \mbox{ :- } B$, 
if $B$ evaluates to $\false$ in $S$ then,
from the definition of $S$, $B$ evaluates to $\false$ in $W$.
(If $b \in B$ is $\false$ in $S$ and $s(b)=+1$, then $not~ b \in W$;
if $not~b \in B$ is $\false$ in $S$ $s(b)=-1$, then $b \in W$.)
It follows that $A$ is unfounded wrt $W$.
Consequently, $not~ a \in W$, which contradicts $a \in S$.
Hence, $A = \emptyset$.

Thus, by the characterization above, $S$ is stable.
\end{proof}
}

As observed above, a second stable model can be obtained by using $\bar{s}$,
if $W$ is not stable.
Along similar lines, but not using signings, \cite{Gire92,Gire94} showed that,
supposing $P$ is order-consistent, if $W$ is not stable then $P$ has at least two stable models.
($P$ is order-consistent \cite{Fages94} if $\lpm$ is well-founded, where
$p \lpm q$ iff $p \lep q$ and $p \len q$.)

Using the construction in the previous theorem, based on Dung's characterization of stable models \cite{Dung92},
we can give a shorter and more direct proof of a variant of Theorem 5.11 of \cite{Dung92} than in that paper.
The proof still uses the same idea as Dung.
The statement of this theorem uses a signing, rather than strictness in \cite{Dung92},
but these two concepts are very closely related (see Section \ref{sec:syntax}).

The \emph{sceptical stable semantics} extending an interpretation $I$
is the set of all literals true in every stable model extending $I$.

\removebrackets
\begin{theorem}[\cite{Dung92}]    \label{thm:Dung}
Let $P$ be a logic program, 
$\P \subseteq \Pi$ be a downward-closed set of predicates with floor $\F$,
and let $\Q$ be  $\Pi\backslash\F$.
Let $I$ be a fixed stable model for $\F$.
Let $W$ be the well-founded model of $P$ extending $I$, 
and $T$ be the sceptical stable semantics of $P$ extending $I$.

If $\Q$ has a signing for $P$ then
$T = W$.
\end{theorem}
\skipit{
\begin{proof}
Every stable model extends the well-founded model, so $W \subseteq T$.
Suppose, to obtain a contradiction, 
$\Q$ has a signing $s$ for $P$ and there is a literal $L$ in $T \backslash W$.
Let the predicate of $L$ be $p$.
Since the semantics of $\F$ is common to $T$ and $W$, $p \notin \F$.

By Theorem~\ref{thm:Turner}, there is a stable model $S$ extending $I$ of $P$ corresponding to $s$,
and another $S'$ corresponding to $\bar{s}$.
Since $L \notin W$, by the construction in that theorem either $L \notin S$ or $L \notin S'$.
But this contradicts the original supposition that $L$ appears in every stable model.
Thus, there is no such $L$ and, hence, $T = W$.
\end{proof}
}

\end{document}